\documentclass[12pt,aps,prb,preprint]{revtex4}   % style for Physical Review B and AJP are similar

\usepackage{dsfont}
\usepackage{amsmath}
\usepackage{graphicx,graphics}
\usepackage{subfigure}

\def\be{\begin{equation}}
\def\ee{\end{equation}}

\def\ba{\begin{array}}
\def\ea{\end{array}}
\def\bea{\begin{eqnarray}}
\def\eea{\end{eqnarray}}
\def\h{\hbox}

\begin{document}

\title{Effects of  curvature on dynamics}
\author{Gautam Dutta}
 \affiliation{Dhirubhai Ambani Institute of Information and Communication Technology,\\
     Gandhinagar 380007, India.}
 \email{gautam_dutta@daiict.ac.in}   %optional
\date{\today}
\begin{abstract}
In this article we discuss the effect of curvature on dynamics when
a physical system moves adiabatically in a curved space. These
effects give a way to measure the curvature of the space
intrinsically without referring to higher dimensional space. Two
interesting examples, the Foucault Pendulum and the perihelion shift
of planetary orbits, are presented in a simple geometric way. A
paper model is presented to see the perihelion shift.

\end{abstract}
\maketitle
\section{Introduction}
How do we know whether the space in which we live is curved or flat?
When we imagine a curved line or a curved surface, like the surface
of a sphere, the inherent picture is the line or the sphere kept in
a flat space of higher dimension. We can then describe the curvature
of these objects in terms of a flat euclidian co-ordinate system.
Imagine two dimensional creatures on the surface of the earth with
no access to height and depth. How do they know they are on the
surface of a sphere and not on a piece of infinite flat land.
 Can an ant know whether the wire it is crawling on is
straight or curved?

The above situations are hypothetical. But how do we answer if the
same question is asked about the three dimensional space we live in.

There is an intrinsic notion of curvature which can be measured
without referring to the external higher dimensional coordinates
\cite{wald}. This is given by the Gauss-Bonet theorem
\cite{henderson} which describes curvature of a surface through
geometrical properties only within the surface. This is like the
curl of a vector field. Though we can express it in terms of a
coordinate system we can define it in terms of the  Stoke's Theorem
without referring to any coordinate system. The Stoke's theorem uses
only the value of a vector field along a closed curve to measure how
much the vector field curls around a region enclosed by the curve.
There are a number of phenomenon in physics where we can see the
effect of a property over a region, within a given boundary, on an
integral done only over the boundary. When a quantum system is moved
adiabatically along a closed curve $C$ the eigenstates evolve and
return to the same state. It however picks up a phase dependent on
the curve $C$. This is purely a geometrical effect independent of
the rate at which the curve $C$ is traversed. This phase is called
the geometric phase or the Berry's phase \cite{berry}. Though these
phases are ignorable for the complete state of the system they can
be observed through interference effects as in Aharanov-Bohm effect
\cite{aharonov}. A similar effect in optics using polarization
states of light was shown by S. Panchratnam \cite{pancharatnam}. A
comprehensive analysis of such phenomena can be found in
\cite{anandan}

%In a curved space  certain postulates of Euclidian Geometry don't hold.
%So the deviation of Euclidian properties of a surface can be a handle to access
%information about the curvature of the space intrinsically.

In a flat space vectors at different locations can be compared since
we have a global coordinate system. We say two unit vectors at
different locations are same if they are parallel. If we transport
one vector to the location of the other they will coincide. This
action of moving a vector from one location to another is called
parallel transport of a vector. On a curved space when we parallel
transport a vector from one location to another the final vector is
not unique. It depends upon the path along which we do the parallel
transport. When the path is a closed loop the vector finally may not
coincide with itself. The change in the orientation of the vector
depends upon the curvature of surface enclosed by the loop.

Laws of Mechanics govern relationships between dynamical vectors. As
the experimental set up moves in space or a particle moves along its
trajectory, the dynamical vectors will undergo parallel transport.
We analyze the effect of curvature on these vectors and how can such
effects indicate the curvature of space intrinsically.

In the next section we present the description of the process of
parallel transport of a vector in a given space. Then we present two
interesting effects of curvature on dynamics. In section 3 we
discuss how the curvature of the latitude along which a Foccoult
Pendulum travels with the earth, affects its plane of oscillation.
In section 4 we discuss in a very simplified and approximate way how
the curvature of space due to gravity causes the perihelion shifts
of the elliptical orbits of planets around the sun. In section 5 we
present the conclusion.

\section{Parallel Transport}
The fifth postulate of Euclid states that ``Given a straight line
and a point outside it, one and only one line can be drawn parallel
to it". This postulate gives a very well defined notion of parallel
vectors at different locations in space. Given a vector $\vec{\bf
v}$ at one location, vectors at other locations are parallel to it
if they are contained by the unique line parallel to $\vec{\bf v}$.
So at every location we can only have one vector parallel to
$\vec{\bf v}$. Here we are only interested in the direction of the
vectors and hence we will only consider unit vectors. Other vectors
are obtained by just multiplying this unit vectors by real numbers.
On a curved surface the fifth postulate of Euclid don't hold. We
don't have a well defined notion of parallel lines on a curved
surface. If parallel lines are defined as lines on the surface that
don't intersect then from a point outside a line we can either have
several lines parallel to it or no lines parallel to it. The
examples of the two types are a  hyperboloid surface and  an
ellipsoidal surface respectively.

In a flat space performing parallel transport of a vector is
straightforward. We have a global cartesian coordinate system and we
just transport the vector along any path keeping its direction fixed
with respect to this cartesian coordinate system. On a curved
surface we don't have a global cartesian coordinate system since we
can't have a family of parallel lines. We only have local tangent
space and we can have coordinate system and vectors only in this
tangent space. As we move along the curve the tangent plane changes.
How do we compare vectors in two different tangent space? The
tangent space represents the vector space at a point on the curved
space. So, as we move from one point to another, there should be no
relative motion between the curved surface and the tangent space.
Such a movement of the tangent plane over the curved surface is
called rolling without slipping. This is like the perfect rolling of
a wheel on a surface, where the part of the wheel near the point of
contact is always at rest with respect to the surface. The vector
being parallel transported must retain its direction with respect to
the coordinate system on this rolling plane. This process defines
the parallel transport of a vector in a curved space.

The direction of motion at each point on the path is along the
tangent to the path. This is also a vector on the rolling tangent
plane. As we move, this vector may change its orientation with
respect to the coordinate system on the tangent plane. If it doesn't
then the path is called a geodesic on the curved surface. A geodesic
is equivalent to a straight line on a flat surface. In a flat space
frames that move in a straight line with uniform speed are inertial
frames. Frames that move along a curve that is not straight in flat
space are non inertial. Likewise, in a curved space frames that move
along a geodesic with uniform speed are inertial frames while frames
that move along any other path on the curved space are non inertial.

\begin{figure}[h]
\includegraphics[scale=0.8]{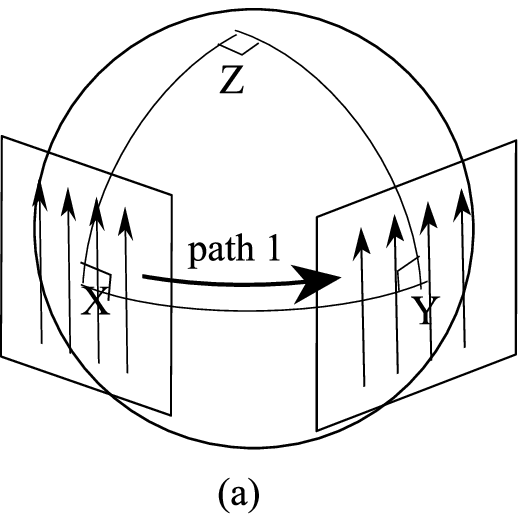}
\hspace{1cm}
\includegraphics[scale=0.8]{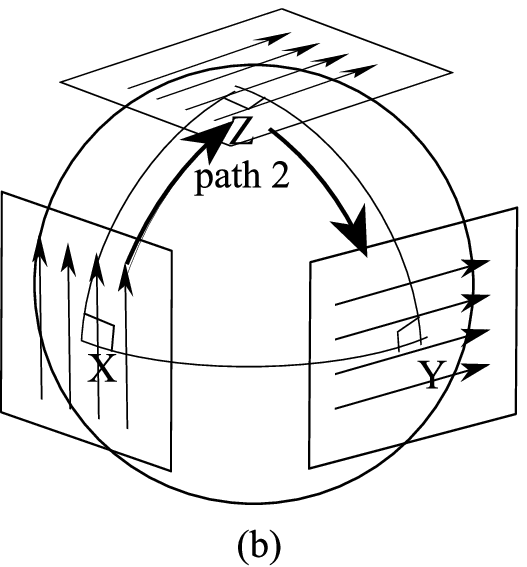}
\hspace{1cm}
\includegraphics[scale=0.8]{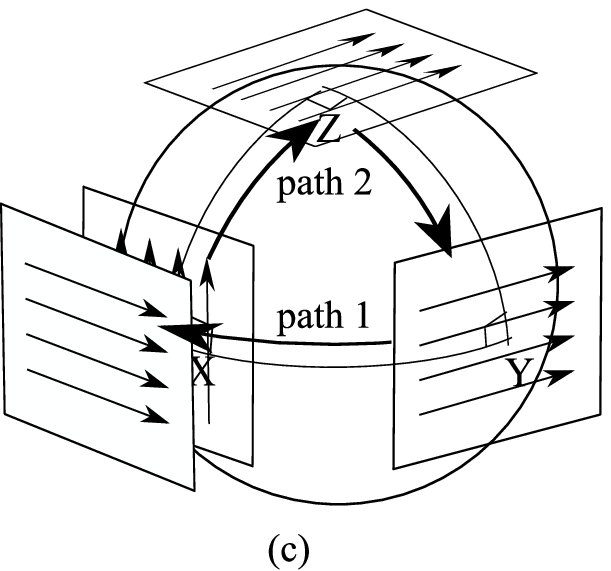}
\caption{X, Y, Z are three points on the sphere connected by
geodesics. (a) and (b) shows
         the tangent plane being parallel transported along two different paths.
         The final orientations of the tangent plane in the two cases are different.
         In (c) the plane is parallel transported along a closed loop. The final orientation of
         the tangent plane is rotated by $90^\circ$ with respect to the initial one.}
         \label{tangentspace}
\end{figure}

Euclidean or flat spaces are characterized by a unique or global
tangent space all over. So whichever path we travel from point X to
point Y, the final tangent space at Y is the same, and it is the
same as the one we had at the point X. If the path is a loop
starting and ending at X, the tangent spaces at the beginning and
the end of the journey will coincide. So every vector parallel
transported along a loop will coincide with itself in a flat space.
In a curved space the tangent vector spaces we obtain after rolling
from point X to point Y depend upon the path taken. This is shown in
Figure~\ref{tangentspace}(a and b) where we roll the tangent plane,
over a spherical surface, along two paths. Path 1 is along a great
circle from X to Y while path 2 is via a third point Z. The paths
from X to Z and Z to Y are great circles which intersect at right
angle at the point Z. We can see that the final orientation of the
tangent plane at Y is different in the two cases. Let us combine the
two paths to form a loop, starting from X, via Z, to Y along path 2,
and coming back from Y to X along path 1. See
Figure~\ref{tangentspace}(c) We can see that the orientation of the
tangent plane at the end of the journey changes from the one we
started with.

The change of orientation of the tangent plane is the same as the
change in orientation of any vector parallel transported along the
curve on the surface since the parallel transported vector always
maintains its direction constant with respect to the co-rolling
tangent planes. So after being parallel transported along a closed
curve a vector coincides with itself on a flat space while it
changes orientation with respect to itself in a curved space. This
change of orientation is related to the curvature over the area of
the surface enclosed by the loop through the Gauss-Bonnet Theorem
\cite{henderson}. If the vector rotates by an angle $\alpha$ as it
goes around the curve $\mathcal{C}$ then 
\be
   2\pi - \alpha = \oint_S {da \over R^2}                            \label{solidangle}
\ee 
where $R$ is the radius of curvature at every point of the
surface and $S$ is the region over the surface enclosed by the
closed curve $\mathcal{C}$. The quantity $da/R^2$ on the r.h.s of
Eq.\ref{solidangle} is the solid angle $d\Omega $ subtended by the
area element $da$ on the surface at the local center of curvature.
The angle $\alpha$ is the amount by which the tangent vector to the
curve rotates on the surface of the manifold as it completes the
loop. So it is a measure of the curvature of the curve within the
manifold, called the intrinsic curvature of the curve.
%The integral on the r.h.s is the
%total solid angle subtended by an equivalent surface $S'$ of a sphere at its center. $S'$ is
%obtained such that the normals at each point of $S'$ are parallel to the normals of $S$.
The curvature at a point on a surface is given as
$\mathcal{K}=1/R^2$, called the gaussian curvature. The solid angle
$\oint_S {\cal K} da$ gives a measure of the total amount by which
the surface curves over the area enclosed by the closed curve
$\mathcal{C}$. In terms of ${\cal K}$ Eq.\ref{solidangle} becomes
\be
   2\pi - \alpha = \oint_S {\cal K} da                             \label{solidangle1}
\ee The Gauss-Bonnet theorem given by Eq.\ref{solidangle1} gives a
way to access the curvature of the manifold, with respect to the
external higher dimensional euclidian space, through the intrinsic
curvature of a curve which can be measured within the manifold.

\section{Foucault Pendulum}
\begin{figure}[h]
\begin{center}
\includegraphics[scale=0.5]{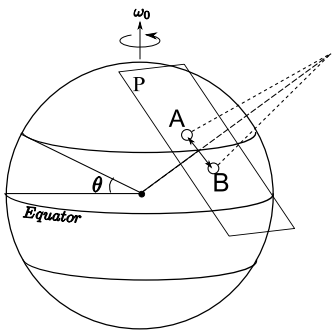}
%\special{psfile="focoult.eps" hscale=50 vscale=50}
\caption{The Foucault Pendulum at a latitude $\theta$ on the earth.
            The bob of the pendulum
         traces a small segment AB on the local tangent plane P to the earth. As the pendulum
         is dragged clockwise along the latitude with the rotating earth, the segment AB is
         parallel transported along the latitude.}
\label{foucault}
\end{center}
\end{figure}

A pendulum performs a simple harmonic motion in the gravitational
field of the earth. A Foucault Pendulum consist of a heavy spherical
bob suspended by a very long thin suspension wire. In the case of
small oscillations the bob of the pendulum lies on the tangent plane
$P$ to the surface of the earth as shown in Figure~\ref{foucault}.

%As the pendulum oscillates
%the suspension wire moves in a plane. This plane is called the plane of
%oscillation of the Pendulum. The difference between an ordinary and Foccoult
%pendulum is that an ordinary pendulum is constrained to oscillate in a
%fixed plane
%whereas the Focoult Pendulum is free to oscillate in any plane passing
%through the local vertical.
The advantage of a foucault pendulum over an ordinary pendulum is
that an ordinary pendulum can oscillate only in a fixed vertical
plane determined by its suspension mechanism. A foucault pendulum
can be set to oscillate along any vertical plane and hence it is not
constrained to change its plane of oscillation along with the
suspension mechanism. The rotation will just cause a torsion in the
thin suspension wire which can get removed by causing a spin in the
bob about the vertical.

%\begin{figure}[H]
%  \begin{center}
%    \includegraphics[height=4in]{focoult.svg}
%    \caption{particle interference}
%  \end{center}
%\end{figure}

As the bob of the pendulum oscillates, it describes a small line
segment AB on the plane $P$. The velocity vector of the bob is along
this line. It is observed that this line segment rotates with
respect to the surface of the earth as the earth rotates. The rate
at which it rotates depends upon the latitude of the location of the
Foucault Pendulum. It is fastest at the pole and slows down as one
moves towards the equator. If $\omega_0$ is the angular velocity of
earth's rotation and $\theta$ is the local latitude then the the
angular speed of rotation of line AB is given by
$\omega=\omega_0\sin\theta$. At the poles this speed is same as that
of the earth's rotation while at the equator it is 0. Also, due to
the factor $\sin\theta$, the sense of rotation is opposite in the
two hemispheres. After one complete rotation of the earth the
segment AB rotates by an angle $2\pi \sin\theta$. The local gravity
everywhere is normal to the plane $P$. As the earth rotates the
pendulum is carried  along a latitude. There is no horizontal force
on the bob to change the direction of its velocity vector, and
hence, that of the orientation of the line AB. This kind of
transport of a vector along a curve is called a parallel transport
of the vector along the curve. The rotation of the plane of
oscillation of the Foucault pendulum is thus the effect of curvature
on the parallel transport of a vector \cite{hart, bergmann, gil}.
The latitude of the earth is not a geodesic. Hence the frame of
reference, fixed with the surface of the earth, carried along the
latitude is not inertial. The rotation of the vector AB in this non
inertial frame is ascribed to a pseudo force called the coriolis
force. So by observing this rotation of AB a two dimensional
creature can only decipher that the surface is rotating with an
angular velocity $\omega_0\sin\theta$. It can't decipher that it is
on a curved surface. However if the creature starts from a point X,
traverses a closed loop, and comes back to the same point, the
parallel transported AB should coincide with its original direction
in a flat space. The change in orientation of AB from its original
direction thus becomes a handle for the two dimensional creature
confined to the surface to measure the curvature of the surface.
\begin{figure}
\begin{center}
%\vskip 5cm \special{psfile="focoult1.eps" hscale=80 vscale=80}
\includegraphics[scale=.8]{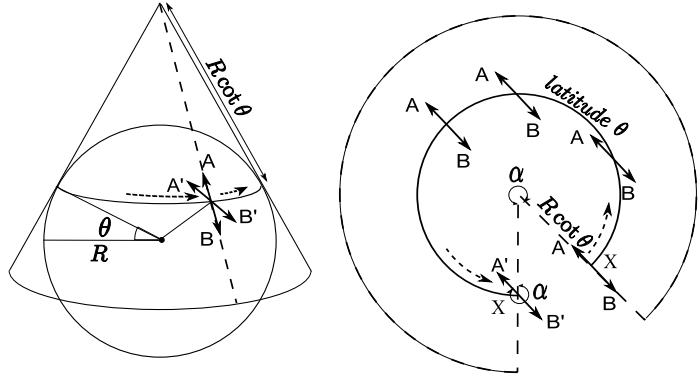}
\caption{ The parallel transport of AB along a latitude
       $\theta$ is equivalent to doing it on the tangent cone touching the earth along the latitude.
       This is shown in the left figure.
       A$'$B$'$ is the final orientation after one complete rotation of the earth.
       The cone is cut
       along the dotted line and shown on the right as a flat sector of a circle of angle $\alpha$.
       AB starts from a point X and moves parallel to itself along the latitude $\theta$.
       It is seen that the angle between AB and the
       latitude gradually changes and finally when it returns to point X the final
       orientation of A$'$B$'$ changes by $\alpha$.}
                                                                           \label{focoult1}
\end{center}
\end{figure}
Let us do the parallel transport of the vector AB along the closed
curve of the latitude of the earth. As the tangent plane rolls along
the latitude the envelope of the tangents is a cone. So it is
equivalent to do a transport of this vector on the surface of this
tangent cone. A cone can be cut and opened to make it flat as shown
in Figure~\ref{focoult1}. The curve of the latitude is shown as the
arc of a circle of radius $R\cot\theta$. Along this arc the segment
AB moves parallelly, and gradually the angle it makes with the local
latitude changes, as shown in the Figure~\ref{focoult1}. Note that
the arc of the circle making the rim of the cone is not complete.
Since the length of the latitude is $2\pi R\cos\theta$ the angle it
subtends at the center is
\[ \alpha= {2\pi R\cos\theta \over R\cot\theta} = 2\pi\sin\theta    \]
It is clear from the Figure~\ref{focoult1} that after completing one
full rotation from point X to the point X, i.e after one day, AB
rotates by an amount $\alpha$ with respect to the local latitude.
According to the Gauss-Bonnet Theorem this change of orientation of
AB suggests a curvature of the surface of the earth enclosed by the
latitude. The total curvature is given by a solid angle measuring
$2\pi -\alpha = 2\pi(1-\sin\theta)$. For a spherical earth this is
precisely equal to the solid angle subtended at its center by the
area enclosed by the latitude $\theta$.

\section{Perihelion Shift}
Planets move around the sun in an elliptic orbit with the sun at the
focus. The nearest point to the sun in the orbit of a planet is
called its perihelion. If the sun is at the origin of the coordinate
system, then the gravitational field in the region around the sun is
given as $k\hat{\h{\bf r}}/r^2$. Bound motion in this gravitational
field can be shown to be elliptic orbits with the sun at the focus.
The energy $E$ and the angular momentum $\vec{\h{\bf L}}$ are the
constants of motion in this field. These constants determine the
shape and size of the orbits, i.e the length of the semi-major axis
and the eccentricity of the ellipse. Since $\vec{\h{\bf L}}$ is a
constant, the orbit of the planet stays in a fixed plane which is
perpendicular to $\vec{\h{\bf L}}$. The orientation of the orbit on
this plane also remains constant. This orientation also depends upon
the initial position and velocity of the planet, just like the
energy and the angular momentum. This orientation is defined by a
vector on the plane of the orbit, called the Laplace Runge Lenz
vector \cite{goldstein} $\vec{\h{\bf A}}$, given by
\[
\vec{\h{\bf A}}=   \vec{\h{\bf p}}\times \vec{\h{\bf L}} - m
{k\vec{\h{\bf r}} \over r}
\]
$\vec{\h{\bf A}}$ is a constant of motion in the Kepler problem. It
can be shown to be directed along the perihelion of the orbit. So
the planetary orbits have a fixed orientation in the plane.

This however fails for mercury, the nearest planet to the sun. It
was known since much before Newton's laws of motion and the theory
of gravity was used to understand planetary orbits. The perihelion
of mercury is not fixed in space, contradicting the constancy of
$\vec{\h{\bf A}}$. Effects due to gravitational pull of venus and
the deviation of the shape of the sun from sphere couldn't explain
this discrepancy. This shift however happens at a very very slow
rate. It is about 43 seconds of an arc in a century. This shift
cannot be understood from Newtonian theory of gravity.

\begin{figure}[h]
\begin{center}
%\vskip 5cm \special{psfile="planet.eps" hscale=80 vscale=80}
\includegraphics[scale=0.8]{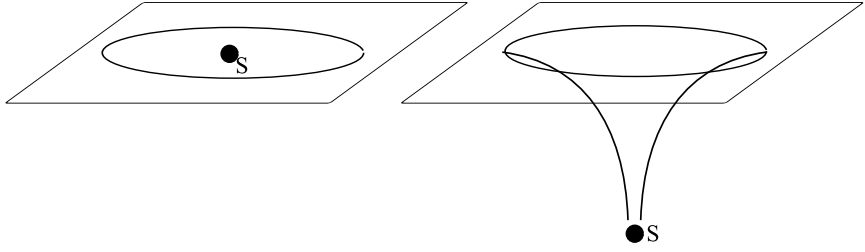}
\caption{Space gets curved near a massive body. The figure on the
left shows a star S and a
         flat plane around it. The figure on the right shows how the plane gets curved near
         the star.}
                                                                           \label{curvedspace}
\end{center}
\end{figure}

In the theory of gravity according to General Relativity \cite{wald,
wheeler}, a massive object causes a curvature of the space-time
around it. All objects in the vicinity of the massive body moves in
this curved space-time. The trajectory is obtained by evaluating the
geodesic path in this space-time. This approach have to be followed
when we are near very massive astronomical bodies like a neutron
star or a black hole. Since the trajectory of planets in a
gravitational field is planar we can consider the effect of
curvature of space-time on this plane. The plane of the trajectory
will now be a curved surface as shown in fig.\ref{curvedspace}.

The curvature of space caused by sun is very small. So the orbit of
mercury is essentially an ellipse perturbed only slightly due to the
curvature of the space.
%This curvature distorts the plane of the orbit of mercury though slightly.
Locally the planet executes an elliptical orbit on the tangent
space. As the planet moves, the tangent plane on the curved space
changes. Any vector which remains constant on the local tangent
space will be parallel transported. Hence the the Laplace Runge Lenz
vector will be parallel transported along the path of the orbit
which lies on the curved space. As we complete the loop on one
revolution we will find that the vector $\vec{\h{\bf A}}$ rotated
with respect to itself. This will cause the perihelion of the
elliptical orbit to shift.
%\begin{figure}[h]
%\begin{center}
%%\hspace{-1cm}
%\includegraphics[scale=0.55]{perihelion-10}
%%\hspace{-2cm}
%\includegraphics[scale=0.55]{perihelion-60}
%\caption{Elliptic orbits on conical surface. The position of the sun
%is at the apex of the cone.
%     In the left figure opening angle of the
%     cone is $\sin^{-1}(59/60)$ while in the right it is $\sin^{-1}(5/6)$. Smaller opening
%     angle indicates high curvature and hence a large perihelion shift.}
%                                                                           \label{perihelion-60}
%     \end{center}
%\end{figure}

\begin{figure}[ht]
\centering \subfigure[]{
\includegraphics[scale=.5]{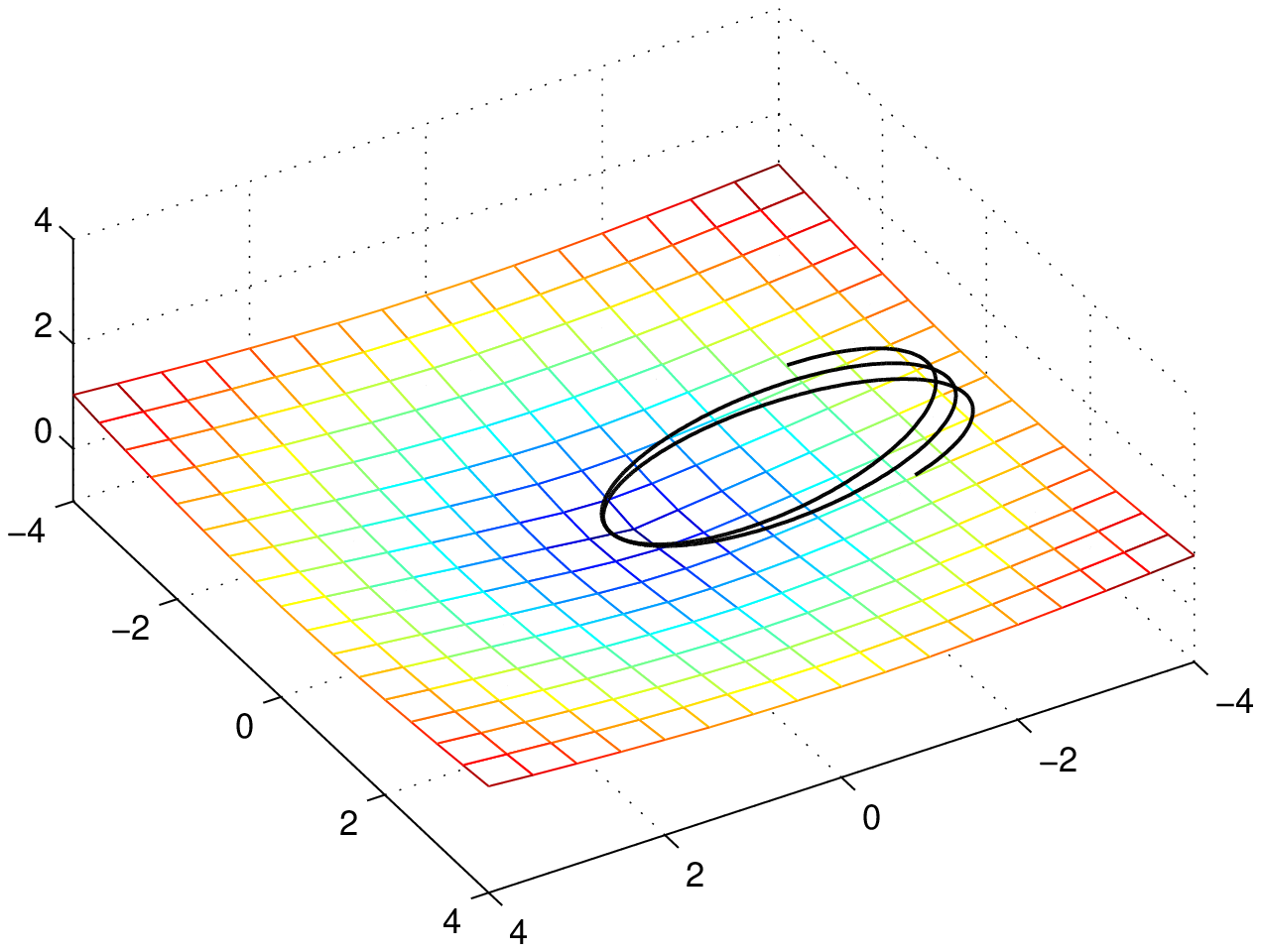}
\label{p1} } \subfigure[]{
\includegraphics[scale=.5]{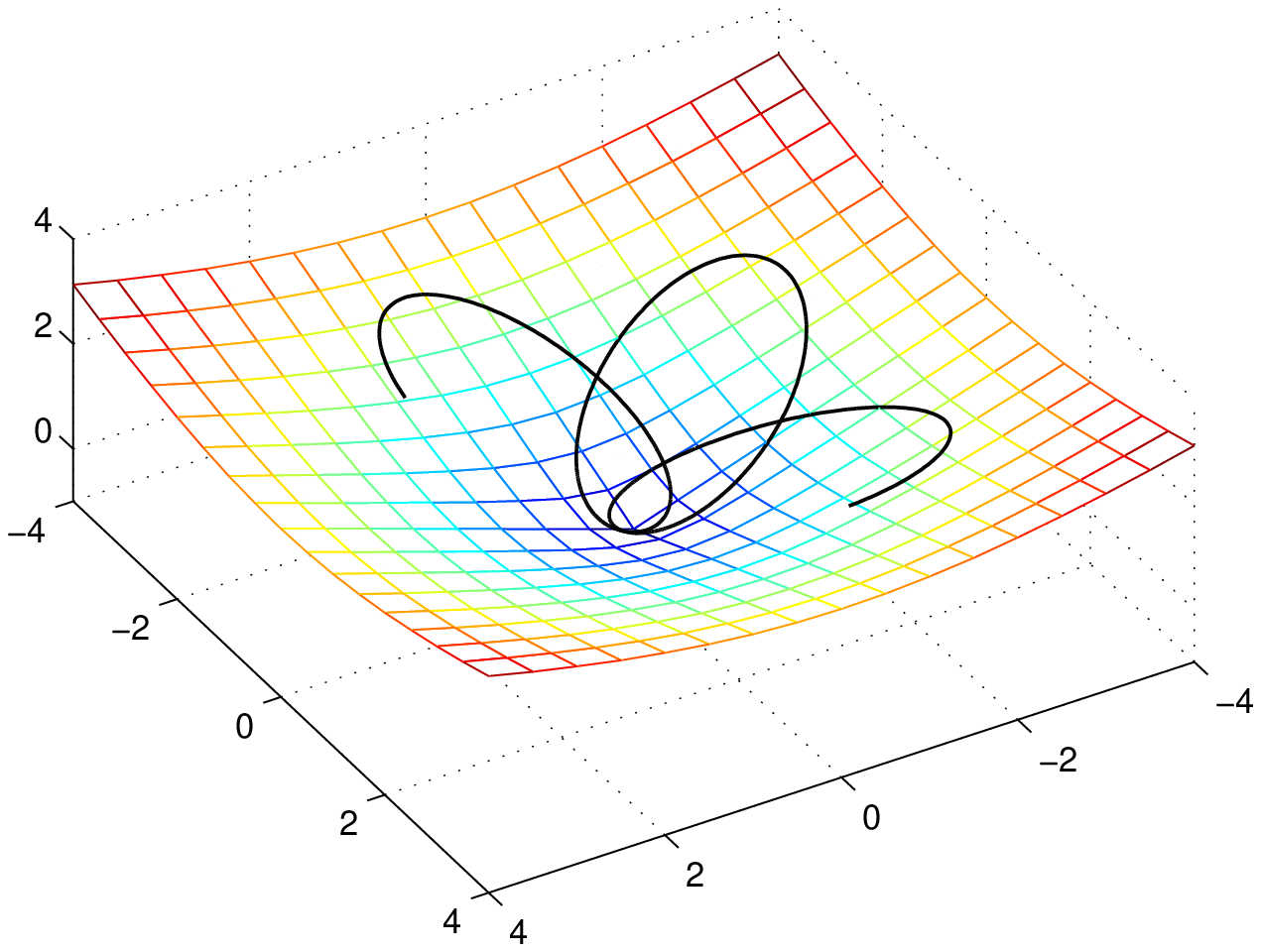}
\label{p2} } \caption[]{Elliptic orbits on conical surface. The
position of the sun is at the apex of the cone.
     In (a) opening angle of the
     cone is $\sin^{-1}(59/60)$ while in (b) it is $\sin^{-1}(5/6)$. Smaller opening
     angle indicates high curvature and hence a large perihelion shift.
\label{perihelion-60}
     }
\end{figure}
We show this with  MATLAB plots in Figure~\ref{perihelion-60}.
Instead of the kind of curved surface as shown in
Figure~\ref{curvedspace}, we work with a cone. This approximation is
fine for orbits of most planets with low eccentricity. The
difference between the nearest and farthest distances in the orbit
from the sun is not much. And so we are essentially on the surface
of a cone. On a flat plane a planet executes a closed elliptic orbit
with the sun at one of the focus. As the plane is curled up into a
cone, with the sun at the apex, the ellipse on the plane is no
longer closed. The perihelion of the ellipse gradually shifts. The
rate at which it shifts depends upon the opening angle of the cone.
When the opening angle is small the curvature due to sun is high. In
Figure~\ref{perihelion-60} we show perihelion shift on two different
conical surface. One with an opening angle of $\sin^{-1}({59\over 60})$
has small perihelion shift while the one with a smaller opening
angle of $\sin^{-1}({5\over 6})$ shows higher perihelion shift.

\begin{figure}[h]
\begin{center}
\includegraphics[scale=0.7]{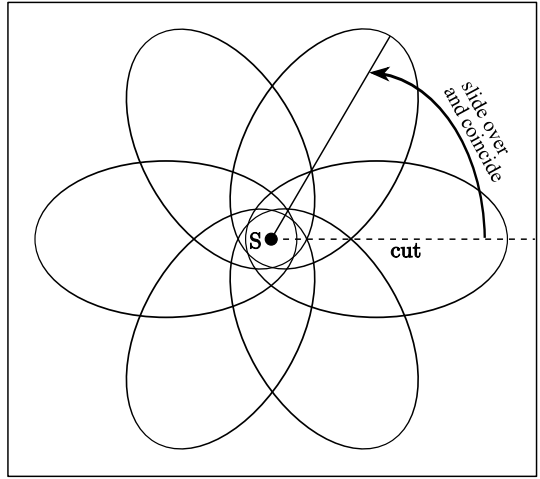}
\caption{A simple paper model to understand the perihelion shift.
After the cone is made trace the ellipse.
         Observe, how the ellipses continuously traverse from one to the other.}
                                                            \label{papermodel}
\end{center}
\end{figure}
A simple paper model can be used to demonstrate the perihelion shift
of elliptic orbits. This is shown in Figure~\ref{papermodel}. We
draw 6 similar elliptic orbits around S as focus with the semi-major
axes of successive ellipses rotated by $\pi/3$. Now we will show how
the orbit smoothly goes from one orbit to the other when this flat
surface is made into a cone. For this cut the paper along the major
axis of one of the ellipse from the edge of the paper to the point
S. Now slide one of the cut edge over the other till one half of the
cut orbit exactly coincides with the other half of the next orbit.
In doing this the plane paper curls up into a cone. The position of
the Sun is at the apex of the cone. Now trace the orbits. The orbits
will smoothly traverse from one to the other. This will give exactly
the effect of perihelion shift of the orbit. Of course the actual
case of mercury has very little shift in one revolution. This is
because the curvature is very very small. To show such small effects
in our simple model we have to draw a large number of ellipses with
their major axes rotated by very small amount and then make a cone
with a very large opening angle. It will be rather difficult to see
the effect. So we are just demonstrating  a rather drastic effect by
a large curvature.

\section{Conclusion}
Curvature of a surface can be generally described in a higher
dimensional euclidian space. But in the absence of any higher
dimensional quantity or the existence of any higher dimensions,
curvature has to be described and measured strictly within the
surface. Gauss-Bonet theorem gives a way to do this. A dynamical
vector that remains constant as the system moves through space acts
naturally as a vector which is parallel transported along a closed
path. The change in the orientation of the vector with respect to
its original orientation after traversing a closed path is a measure
of the total curvature of a surface enclosed by the path. The
foucault pendulum and the perihelion shift of the orbit of a planet
are two such examples. While foucault pendulum indicates the
curvature of the surface of the earth embedded in higher three
dimensional space, the perihelion shift indicates the curvature of
the space itself which can't be described in any higher dimensional
manifold. Very simple geometric demonstration of these effects are
presented.


\begin{thebibliography}{5}
\bibitem{wald} Robert. M. Wald, {\it General Relativity}. (The University of Chicago Press,
               Chicago and London 1984).
\bibitem{henderson} David W. Hnderson, {\it Differential Geometry, A Geometric Introduction}.
               (Prentice Hall. 1998)
\bibitem{berry} M. V. Berry, {\it Proc. R. Soc. Lond.} A, {\bf 392}, 45-47 (1984).
\bibitem{aharonov} Y. Aharonov, D. Bohm, ``Significance of electromagnetic potentials in the
               quantum theory", Phys. Rev. {\bf 115} 485-491 (1959).
\bibitem{pancharatnam} S. Pancharatnam, ``Generalized theory of interference, and its applications",
               Proc. Indian Acad. Sci. A {\bf 44} 247-262 (1956)
\bibitem{anandan} J. Anandan, J. Christian, K. Wanelik,
                ``Resource letter GPP-1: Geometric phases in Physics", Am. J. Phys. {bf 65}(3),
                180-185 (1996).
\bibitem{hart} J. B. Hart, R. E. Miller, R. L. Mills, ``A simple geometric model for visualizing the
                motion of a Foucault pendulum", Am. J. Phys. {\bf 55}(1), 67-70 (1987).
\bibitem{bergmann} J. von Bergmann, Hsing Chi von Bergmann, ``Foucault pendulum through basic geometry",
                Am. J. Phys. {\bf 75}(10), 888-892 (2007).
\bibitem{gil} S. Gil, ``A mechanical device to study geometric phases and curvatures",
                Am. J. Phys. {\bf 78}(4), 384-390 (2010).
\bibitem{goldstein} Herbert Goldstein, {\it Classical Mechanics}. (Addison Wesley Publishng Company,
               Inc., USA 1980)
\bibitem{wheeler} C. W. Misner, K. S. Thorne, J. A. Wheeler, {\it Gravittion}.
               (Freeman, San Francisco 1973)
\end{thebibliography}
\end{document}